\documentstyle[conf-X]{article}
\begin{document} 
\small
\heading{%
%
X-rays and Cosmology
}
\par\medskip\noindent
\author{%
A.C. Fabian$^1$
}
\address{%
Institute of Astronomy, Madingley Road, Cambridge CB3 0HA, UK
}
%

\begin{abstract}
The role of X-ray observations for cosmological studies and
conclusions is briefly explored. X-rays currently yield cosmologically
interesting results on the abundances, evolution and gas content of
clusters of galaxies, on the clustering and evolution of active
galaxies and on the X-ray Background. They are unlikely in the long
term future to give the most precise values of the cosmological
parameters, although in the short term the baryon fraction of clusters
and the Sunyaev-Zeldovich effect will remain important determinants
and checks for some parameters. X-rays will however continue to play
an important role in studying the astrophysics of the formation and
growth of black holes, galaxies, groups and clusters. It is possible
that this role will be crucial, if winds from active galaxies are
responsible for breaking the simple gravitational scaling laws for
clusters.
\end{abstract}

\section{Introduction}

X-ray observations have in the past played a minor, but important
role, in cosmology. For example, the hot intergalactic medium
prediction of the Steady State theory failed because the observed
intensity was too low (Gould \& Burbidge 1963). More recently,
estimates of the gas fraction in clusters, which is dominated by the
X-ray emitting intracluster medium, have shown that the matter density
of the Universe is significantly below the critical value for closure
(S White et al 1993; D White et al 1995).

X-rays may not be the best means to quantify the cosmological
parameters ($H_0, \Omega_{\rm m}, q_0, \Lambda$ etc). Of course we
should continue to do our best, but it is unlikely that X-ray results
will ever supercede those expected from anticipated studies of the
Cosmic Microwave Background. At present they supplement such work
(e.g. Bridle et al 1999; Bahcall et al 1999).

Cosmology for many of us is not, though, just the study of the
geometry of the Universe, or of the inflation era, but involves the
study of what has happened since redshifts of, say, 10. It involves
the {\it astrophysics} of the baryons in this recent era, and working
out how stars, black holes, galaxies and clusters arose and why. From
that point of view, X-rays will be vital, since much of the
dissipation in galaxy formation occurs at (soft) X-ray emitting
temperatures; the X-ray emitting intracluster and intragroup medium is
a living fossil, or calorimeter of the enrichment and energy injection
history of galaxies; the central black holes in galaxies may only be
clearly studied from their penetrating X-ray emission, and the
isotropy of the X-ray Background may be guide to the development of
structure. The deepest, and the most extended, gravitational potential
wells in the Universe are potent X-ray emitters.

I shall not in this introductory talk discuss the possible X-ray
detection of most of the baryons in the local Universe through the
study of groups, since this is covered by R Mushotzky; nor shall I
explore the implications of the fluctuations in the X-ray Background,
which are covered by X Barcons. Mainly I shall discuss some issues
which I have been working on and which illustrate the main theme of
this Conference.

\section{Clusters and Cosmology}

Clusters are the most massive virialized objects in the Universe and
so their number density is a key diagnostic of the cosmic power
spectrum, on a scale of about 8~Mpc, $\sigma_8$. The evolution of
their numbers, and particularly of the numbers above a particular
intracluster gas (virial) temperature, is a good guide to the mean
matter density in the Universe, $\Omega_{\rm m}$ (Eke et al 1998).
Current work suggests that if there is any evolution out to redshifts
of about one, then it is very small (Ebeling 1999). The existence of
the X-ray luminous EMSS cluster 1054-0321 provides a strong argument
that $\Omega_{\rm m}$ is significantly less than one (Gioia et al
1998).

Together with radio or microwave observations, X-ray intensities yield
interesting constraints on the Hubble constant through the S-Z effect.
The common occurrence of cooling flows at the centres of clusters, and
recent working on ageing the flows (Allen et al 1999), indicates the
merger history of cluster cores and gives a clue to possible baryonic
dark matter.

Perhaps the most striking application of X-rays has been through the
baryon fraction, $$f_{\rm b}={{\rm mass\ of\ baryons}\over{\rm total\
mass}}={{\Omega_{\rm b}}\over{\Omega_{\rm m}}}.$$ Since $\Omega_{\rm
b}$ is known fairly well from the combination of cosmic
nucleosynthesis and deuterium abundance work, $\Omega_{\rm m}$ is well
determined from X-ray cluster studies (White \& Frenk 1991). The value
found is about 0.3.

Many clusters have now been carefully observed and their gas fractions
determined out to $r_{500}$ or more. Specifying a radius is important
as the fraction appears to rise with radius, from about 10 per cent in
the core to about 20 per cent at $r_{500}$, within which the mean
cluster density is 500 times that of the Universe. In a sample studied
by Ettori and myself (1999), we find that the gas fraction drops with
redshift as $(1+z)^{1.5}$, which is surprisingly steep given the lack
of evolution seen in other properties. It is explained by our use of
$q_0=0.5$ when determining $r_{500}$ (Sasaki 1996: for an isothermal
cluster the gas mass $M_{\rm g}\propto n r^3$, which since $L\propto
n^2 r^3$ means that $M_{\rm g}\propto L^{1/2}r^{3/2}\propto D_{\rm
A}$, where $D_{\rm A}$ is the angular diameter distance of the
cluster, then since the total mass $M_{\rm T}\propto T r \propto
D_{\rm A},$ we have $M_{\rm g}/M_{\rm T} \propto D_{\rm A}^{3/2}$ ).
If we adjust $q_0$ (thus $D_{\rm A}$) so that the gas fraction is
constant with redshift, then we find that $q_0<0.1$, in other words it
provides a strong case for a low matter density Universe ($\Omega_{\rm
m}<0.2$). Why it gives a lower value than that found via cosmic
nucleosynthesis is not clear, however.

A further cosmologically interesting result from cluster studies with
X-rays is that of scaling, particularly for the luminosity --
temperature relation. From gravitational collapse alone, we expect
that $L_{\rm x}\propto T_{\rm x}^2$ (the virial radius $r_{\rm v}
\sim r_{200}$, defined as for $r_{500}$ above, within which
the mean density of all clusters is the same at the formation
redshift, therefore $M r_{\rm v}^{-3} = {\rm const}$ and since $T_{\rm
x}\propto M r_{\rm v}^{-1}$ from the virial theorem, then $T_{\rm
x}\propto r_{\rm v}^2$; then since $L_{\rm x}\propto n^2 T_{\rm
x}^{1/2} V$ and the mean density of clusters $n$ is constant, then
$L_{\rm x}\propto T_{\rm x}^2)$. The observations however indicate
$T_{\rm x}^3$ over the range $kT_{\rm x}\sim 2-8$~keV. At higher
temperatures it appears to level off more to $T_{\rm x}^2$ (Allen \&
Fabian 1998) and is steeper at lower temperatures.

A major implication of this departure from the expected scaling is
that significant amounts of heat have been injected into clusters. Wu,
Nulsen and I (1999) find it difficult to explain this in terms of
supernova heating (but see Loewenstein 1999). Of course supernovae
have enriched the intracluster medium in metals, but is likely that
much of their heat was radiated from the interstellar medium of their
host galaxies, If it was not then supernovae can hardly provide the
feedback required for galaxy formation to result in the galaxies seen
today rather than many small dense ones. We also investigated
widespread cooling as a way of removing low entropy gas from cluster
cores and also preheating to give an entropy floor to cluster gas
Ponman et al 1999). None appears to be sufficient.

We also investigated the limits that the $1/4$~keV intensity of the
X-ray Background gives to galaxy formation (Wu, Fabian \& Nulsen 1999;
see also Pen 1997). The intensity predicted without any heat source
apart from gravity is about an order of magnitude greater than that
observed. Again a significant heat source is required, this time in
all objects exceeding a few times $10^{12}$ solar masses. The total
heat requirement to solve both the soft X-ray background problem and
the scaling one is about 3~keV per particle, which is high.

A possible and perhaps plausible heat source is winds from active
galaxies, an issue I return to near the end. If a significant fraction
of the power expected from the formation of the local mass density of
massive black holes emerged in winds, as well as radiation, then the
problems are solved. 

An implication of this is that the intergalactic medium may also have
this mean energy, either in heat or potential energy, and so be much
more difficult to detect then predicted by Cen \& Ostriker (1999).

\section{Active Galaxies and the X-ray Background}

The X-ray Background (XRB) is the sum of all the X-ray emission in the
recent Universe. Since the spectrum in the 2--10~keV band is flatter
than known classes of source, it is most probable that it is the sum
of many obscured active galaxictic nuclei (AGN). This explanation
arose over ten years ago (Setti \& Woltjer 1989) and has been explored
in detail since (Madau, Ghisellini \& Fabian 1994; Celotti et al 1995;
Matt \& Fabian 1994; Comastri et al 1995; Wilman \& Fabian 1999). A
simple comarison of the XRB spectrum with that of an unobscured AGN,
with a typical photon index of 2, demonstrates that {\it most
accretion in the Universe is obscured} (Fabian et al 1998; Fabian \&
Iwasawa 1999).

A robust estimate of the accretion power in the Universe, assumed to
be from AGN, can be obtained by assuming that the intensity of the XRB
at 30~keV is emitted by an underlying power-law and yet least affected
by photoelectric absorption (Fabian \& Iwasawa 1999). Matching that to
the spectral energy distribution of an unobscured AGN (see e.g. Elvis
et al 1994), then allows the absorption-corrected energy density from
accretion to be determined, $\varepsilon_{\rm AGN}$. It can be
increased by a correction for Compton thick objects (where $N_{\rm
H}>1.5\times 10^{24}$), by a factor of 1.3 (Maiolino et al 1998) to
perhaps as much as 2. Then Soltan's (1982) argument can be used to
convert to an expected mean mass density in black holes now;
$$\rho_{\rm BH}={{{(1+\bar z)}
\varepsilon_{\rm AGN}}\over {0.1 c^2}},$$ where $z\sim2$ is the mean
redshift of the AGN, and an accretion efficiency of $0.1$ has been
assumed. The result is about half the mass density found by Magorrian
et al (1998) and is in good agreement with van der Marel's (1999)
value (see also Salucci et al 1999). It means that most of the mass of
black holes is due to radiatively efficient, but obscured, accretion.
Most, about 85 per cent, of the accretion power has been absorbed and
presumably reradiated in the infrared. The total radiated power from
AGN can be seen from the IR backgrounds (e.g. Fixsen et al 1998) to
then be about one quarter (give or take a factor of two) of that from
stars.

Note here that X-rays are the only radiation which penetrates the
absorber directly and so can discriminate and inform us of the actual
evolution of AGN and their black holes. The radiation is absorbed at
other wavelengths and so provides no direct information on the central
source. It is likely that this also applies to the cores of most
galaxies, which are the oldest parts. X-ray observations are likely to
be crucial to understanding the accretion history of the Universe and
of the evolution of the dense cores of galaxies. If the black holes in
them formed very early and have always been obscured, then X-rays may
become important for studying the earliest objects.

There may be an intimate connection between the formation of a galaxy
spheroid and its central black hole (Silk \& Rees 1998; Fabian 1999).
If part of the gas forming the spheroid remains as cold obscuring gas
clouds, instead of rapidly forming stars, and if the central black
hole grows by accretion and blows a wind, then it may blow away the
gas, and end the growth of both the black hole and the spheroid when
it becomes massive and powerful enough. This scenario means that the
main growth phase of the black hole is obscured, and it only becomes
unobscured when the fuel supply is blown away. It then lasts for a
disk emptying time, which may be say 10 per cent of its original
lifetime. As the gas is blown away it may be seen as a broad
absorption line quasar. 

The implication here is that AGN have powerful winds, especially when
they accrete near the Eddington limit. We have seen from the cluster
discussion that such powerful winds can provide and explanation for
the excess energy in clusters. Thus the total power of AGN and the
excess heat in clusters may be intimately linked in this way.

\section{Conclusions}

X-ray observations enable the thermal content and enrichment of the
baryons in the Universe to be studied. The cooling of gas in the
potential wells of galaxies, groups and clusters also emits
predominantly in X-rays. They are the best direct probe of the
accretion history of AGN, the integrated energy of which may be up to
50 per cent of that from stars.

\acknowledgements{I am grateful to the organisers of this
meeting for the opportunity to visit Santorini and hear the many
interesting talks and discussions, as well as to my collaborators
Stefano Ettori, Kazushi Iwasawa, Paul Nulsen and Kelvin Wu. The Royal
Society is thanked for support.}

\vfill
\end{document}